\documentclass[twocolumn]{revtex4}
\usepackage{amssymb,amsmath}
\usepackage[dvips]{graphicx}

\newcommand{\be}{\begin{equation}}
\newcommand{\ee}{\end{equation}}
\newcommand{\ud}{\mathrm{d}}

\begin{document}

\title{
    \vskip -120pt
    {
        \begin{small}
        \hfill DAMTP-2008-39   \\
        \end{small}
    }
    {\bf\large
        Primordial black holes in the Dark Ages: \\
	Observational prospects for future 21cm surveys
    }
}

\author{Katherine J. Mack}
\email{mack@astro.princeton.edu}
\affiliation{Department of Astrophysical Sciences, Peyton Hall\\
Princeton University, Princeton, NJ 08544 USA}
\author{Daniel H. Wesley}
\email{D.H.Wesley@damtp.cam.ac.uk}
\affiliation{Centre for Theoretical Cosmology 
DAMTP, Cambridge University\\ 
Wilberforce Road, Cambridge, CB3 0WA United Kingdom}

\begin{abstract}
\noindent
We consider the signatures of a population of primordial black holes (PBHs) in future observations of 21cm radiation from neutral hydrogen at high redshift.  
We focus on PBHs in the mass range $5 \times 10^{10} \textrm{kg} \lesssim M_{\textrm{PBH}} \lesssim 10^{14} \textrm{kg}$, which primarily influence the intergalactic medium (IGM) by heating from direct Hawking radiation. 
Our computation takes into account the black hole graybody factors and the detailed energy dependence of photon and $e^\pm$ absorption by the IGM.
We find that for black holes with initial masses between $5 \times 10^{11} \textrm{kg} \lesssim M_{\textrm{PBH}} \lesssim 10^{14} \textrm{kg}$,
the signal mimics that of a decaying dark matter species.
For black holes in the range $5 \times 10^{10} \textrm{kg} \lesssim M_{\textrm{PBH}} \lesssim 5 \times 10^{11} \textrm{kg}$, the late stages of evaporation produce a characteristic feature in the 21cm brightness temperature that provides a unique signature of the black hole population.  
If no signal is observed, then 21cm observations will provide significantly better constraints on PBHs in the mass range $5 \times 10^{10} \textrm{kg} \lesssim M_{\textrm{PBH}} \lesssim 10^{12} \textrm{kg}$ than are currently available from the diffuse $\gamma$-ray background.
\end{abstract}

\maketitle
\newpage

\section{Introduction}

The 21cm hyperfine spin-flip transition of neutral hydrogen (HI) may allow observers to probe the cosmic ``Dark Ages" comprising the epoch between the last scattering of the cosmic microwave background (CMB)  at 
$z\sim 1100$, and the appearance of luminous sources at $z\sim 30$
\cite{Barkana:2006ep,Furlanetto:2006jb,Pritchard:2008da}.  Observations at wavelengths $21(1+z)$ cm could be used to slice the universe as a function of redshift $z$, and so produce a three-dimensional map of the HI distribution when combined with angular measurements. At moderate redshifts, this could allow reionization of the universe to be studied in detail. Overall, the essentially three-dimensional nature of 21cm data gives it the potential to become one of the the richest data sets for cosmology \cite{Loeb:2003ya}.  It will not only enable the standard $\Lambda$CDM cosmological model to be better tested and understood, but will provide a new means to constrain -- or detect -- more exotic possibilities.  One such possibility, the presence of a population of primordial black holes (PBHs), is ideally suited to study with the 21cm background, as we describe in this work.

The 21cm signal is sensitive to exotic physics through its dependence on the thermal history of the intergalactic medium (IGM).  The IGM is visible in 21cm when the spin temperature $T_S$ of the neutral hydrogen gas differs from the CMB temperature $T_{\textrm{CMB}}$.  The spin temperature $T_S$ is itself determined by the competition between its coupling to $T_{\textrm{CMB}}$ through interactions with CMB photons, its coupling to the gas kinetic temperature $T_K$ through atomic collisions, and (especially at lower redshifts) interactions with Ly-$\alpha$ photons produced by luminous sources.  Any process that heats the IGM will influence $T_K$ and therefore affect $T_S$.  For redshifts $z\sim 30-300$, one expects $T_K \sim 10-10^3 K \sim 10^{-3}-10^{-1}$ eV, and so little heating per hydrogen atom is necessary to significantly change the thermal history of the IGM and influence the 21 cm signal. By contrast, the CMB is mainly sensitive to the Dark Age IGM thermal history through the effects of IGM heating on the ionized fraction, which affects the optical depth to the last scattering surface $\tau_{\textrm{LSS}}$.  The CMB is therefore less sensitive to IGM heating, since the energy required to ionize HI is much larger than the typical $T_K$ over the redshift range of interest.  Furthermore, changes in the CMB TT power spectrum due to  $\tau_{\textrm{LSS}}$ are degenerate with changes due to the scalar spectral index $n_s$ (though measurements of the TE power spectrum can break this degeneracy).  The projected sensitivity of 21cm emission to the IGM thermal history has been exploited to show that the 21cm signal can provide much more stringent constraints on a population of long-lived, decaying particles than is available with CMB data \cite{Furlanetto:2006wp}.

Primordial black holes (PBHs) are an excellent exotic physics target for 21cm observations because the mechanisms by which they heat the IGM are under tight theoretical control.  In this work, we compute the effects of a population of PBHs with masses between $10^{10}$ and $10^{15}$ kg on 21cm observables, including the sky-averaged brightness temperature and the fluctuation power spectrum.  We also obtain predictions for the relic photon density from the population of PBHs.  We take into account the details of photon and $e^\pm$ emission by PBHs and the energy-dependent absorption properties of the IGM.   We find that future 21cm observations can provide better constraints on PBHs than are currently available.  The techniques and computer code we develop can be easily modified to study other exotic physics scenarios in the Dark Ages, such as dark matter annihilation and decay, if the precise energy spectrum of photons and $e^\pm$ pairs produced by the exotic physics mechanism is known.

The effect of PBHs on the IGM is strongly dependent on the PBH mass.  Very massive PBHs would accrete matter in accretion disks and thereby emit x-rays; this scenario is explored in \cite{Ricotti:2007au}.  We focus our present discussion on low-mass PBHs that would primarily affect the IGM through the Hawking radiation \cite{Hawking:1974rv,Hawking:1974sw} of light particles: primarily photons and $e^\pm$ pairs, and heavier particles (and their decay products) for the smallest holes.  The total power emission by black holes through these channels has been known for some time as a function of the black hole mass~\cite{Page:1976df}.  Due to a coincidence between the Hawking temperature of PBHs that evaporate during the Dark Ages and a window of low optical thickness of the IGM to photon absorption,
it is vitally important to know the spectrum of photons produced by the PBHs.  The radiated photons have energies such that the probability of absorption by the IGM, and thus their effectiveness in heating it, is strongly dependent on the photon wavelength.  The emission rates as a function of photon energy are given by the black hole ``graybody factors," which have been calculated.  The graybody factors arise because black holes should not radiate with a blackbody spectrum, even though they act as warm bodies with a specific temperature.  In the end, all that is needed to completely characterize the PBH population is the total density parameter $\Omega_{\textrm{PBH}}$ and the mass of the individual black holes $M_{\textrm{PBH}}$.  By comparison, while in principle it is possible to calculate the spectrum of photons (and other particles) produced in dark matter decay or annihilation, in practice this dependence is subsumed in an overall energy deposition rate per baryon $\epsilon$.  There is then a model-dependent conversion between $\epsilon$ and quantities such as the density parameter and lifetime of the relevant particle species.  For PBHs, the energy deposition rate is uniquely determined once the mass of the black holes is specified. 

We obtain our predictions as follows. We numerically integrate the equations governing IGM ionization and temperature, including exotic sources of energy injection.  We use a modified version of the RECFAST \cite{Seager:1999bc} code at high redshift for increased accuracy, and a simpler set of IGM evolution equations at lower redshifts. The energy injection itself is computed by using the graybody factors for Hawking emission and the optical depths for a variety of IGM photon and $e^\pm$ absorption processes.  Some of these processes redistribute photons by energy, and so we track the full photon population as a function of energy through cosmic history.  We use this information to compute the IGM temperature and ionization history, and  use standard techniques to compute the 21cm brightness temperature and the fluctuations.  In the end, we obtain the complete temperature and ionization history of the IGM, the power spectrum and sky-averaged signal of 21cm brightness temperature fluctuations, and a relic high-energy photon population.  In order to obtain a constraint on PBHs, we compare the power spectra from several models to the power spectrum with no PBHs included and use an estimate of the measurement error for a realistic future experiment.

The PBH population is currently most tightly constrained for PBH masses near $10^{12}$ kg, by the diffuse $\gamma$-ray background as measured by EGRET \cite{Sreekumar:1997un,Carr:1998fw}.  The highest-energy EGRET photons cannot be of primordial origin ($z > 10^3$) because the IGM is not transparent at the corresponding energies.  These photons are thought to originate from outside the Galaxy, but their production mechanism is as yet unknown.
One possibility is that these high-energy photons arise from processes at redshifts lower than that of reionization, in which case it is unlikely that 21cm observations will be a useful probe.  Another possibility is that these photons arise from some energetic process occurring during the Dark Ages.  In this case the details of photon absorption and heating by the IGM will be important.  If these energetic processes heat the IGM, then 21cm observations could provide a complimentary measurement and a consistency check on exotic physics mechanisms which purport to contribute to the high-energy diffuse $\gamma$-ray background.

This paper is organized as follows.  We review the physics of PBH formation and energy emission in Section \ref{s:bh_emission}.  In Section \ref{s:IGMphysics} we describe our computation of the energy injection rates from PBHs and how this influences the mean IGM evolution, the 21cm brightness temperature, and the 21cm power spectrum.  We give an overview of our results in Section \ref{s:numerical}.  We include predictions for the ionization history, 21cm observables, and relic photon population for a selection of PBH models.  In Section \ref{s:results} we present the constraints on PBH populations that can be obtained with an ambitious 21cm experiment.  We discuss observational prospects in Section \ref{s:observation} and conclude in Section \ref{s:discussion}.

\section{Primordial black hole emission} \label{s:bh_emission}

\subsection{Formation mechanisms} \label{s:formation}

Several possible formation mechanisms for PBHs have been discussed in the literature, and the resulting mass spectra are highly model-dependent.  We briefly mention some possible PBH formation mechanisms here.  For a more in-depth review, see, e.g., \cite{Carr:2005bd}.

As first proposed, PBHs result from fluctuations in the high density primordial perturbations that collapse upon horizon entry \cite{Zeldovich:1967,Hawking:1971ei}.  This scenario is expected to produce an extended PBH mass spectrum \cite{Carr:1975qj}. As it generally requires a spectral index $n_s > 1$, this scenario is disfavored by the most recent observations which indicate $n_s < 1$ (e.g., the WMAP five-year data set \cite{Spergel:2006hy,Dunkley:2008ie}).  Other PBH formation scenarios invoke a phase transition or a period of a softening equation of state during which fluctuations would be more likely to collapse (e.g., \cite{Jedamzik:1996mr}). These models predict a very narrow mass spectrum, as PBH formation would occur only at the time of the transition.  Other formation scenarios invoke more exotic mechanisms, such as the collapse of cosmic string loops \cite{Caldwell:1995fu,Garriga:1993gj,Hawking:1987bn,Polnarev:1988dh,MacGibbon:1997pu} and of domain walls \cite{Berezin:1982ur,Ipser:1983db}, and predict a broader mass spectrum.

In this work, we assume all of the PBHs have the same mass, amounting to a delta-function mass spectrum.  This approximation is good for some PBH formation scenarios and less so for others.  It is the most general in the sense that its results may be easily applied to other models.  Our results show that the constraints on the PBH population depend strongly on mass, but in a simple way.  By assuming a single PBH mass, we can determine the mass ranges most tightly constrained by 21cm observations. We expect the constraints on a more general PBH mass spectrum to be determined, to a rough approximation, by the number density of PBHs in the mass range to which 21cm observations are most sensitive.  Therefore assuming a single PBH mass gives a good indication of the constraints on more general PBH mass distributions.

\subsection{Hawking radiation} \label{s:hawking}

The primary means by which black holes with masses $\lesssim 10^{14}$ kg would have influenced the IGM is through Hawking radiation.  It has been known for some time that a Schwarzschild (uncharged, non-rotating) black hole of mass $M_{\textrm{PBH}}$ should radiate \cite{Hawking:1974rv,Hawking:1974sw} as would a warm body at the temperature $T_H$, where
\begin{align}\label{eq:HawkT}
T_H = \frac{\hbar c^3}{8\pi G M_{\textrm{PBH}} k_B} & = \frac{1.2\times 10^{13}\;{\rm K}}{(M_{\textrm{PBH}} / 10^{10} \; {\rm kg})} \notag \\
& = \frac{1.1 \;{\rm GeV} / k_B}{(M_{\textrm{PBH}} / 10^{10} \; {\rm kg})}.
\end{align}
The black hole should emit all massless and nearly-massless particles (gravitons, photons, neutrinos), as well as those massive particles whose mass is substantially below $k_B T_H$.  The energy emitted in neutrinos and gravitons is essentially lost, for these particles interact very weakly and do not affect the IGM.
On the other hand, the IGM is affected by the energy emitted in photons and by the $e^\pm$ pairs emitted by smaller black holes.  

While black holes are expected to have a temperature, they should not emit radiation with a blackbody spectrum. Instead,  Hawking's calculation shows that the number of particles emitted with angular frequency $\omega$ (measured at infinity), spin $s$, polarization $p$, and angular momentum quantum numbers $lm$ is
\be
\langle N_{splm}(\omega) \rangle
= \frac{ \Gamma_{splm}(\omega) }{ \exp\left(\hbar\omega / k_B T_H\right) - (-1)^{2s} },
\ee
where the $\Gamma_{splm}(\omega)$ are  graybody factors. The $\Gamma_{splm}(\omega)$ parameterize the deviation of the black hole emission spectrum from that of a blackbody.  Physically, each $\Gamma_{splm}(\omega)$ is the probability that a particle in an infalling field mode described by $sp lm$ is absorbed by the hole.  Since a black hole with temperature $T_H$ must be in thermal equilibrium with a blackbody heat bath of the same temperature, it follows that these absorption probabilities must also determine the emission rate.  In addition to the parameters $splm$, the $\Gamma_{splm}(\omega)$ also depend on the mass of the particle species.  Only particles with mass $m \ll T_H$ (in units where $k_B=c=1$) will be emitted at an appreciable rate, and for these particles the $\Gamma_{splm}(\omega)$ are those of massless particles.  In this work we only consider the emission of particles with $m \ll T_H$ and so use the massless graybody factors.

It is essential to include the graybody factors when studying the effect of black hole emission on the IGM.  They determine the total power emission in particles of various spin, and therefore the evolution of the black hole mass.  The rate of IGM photon absorption is very frequency-dependent, and so it is essential to know the photon spectrum accurately.  We use the power emission spectra computed in \cite{Page:1976df}, obtained by integrating the 
Press-Teukolsky equations for fields of different spin \cite{Teukolsky:1973ha,Teukolsky:1974yv}.  A general feature of these graybody factors is that the emission of particles of high spin is suppressed, and the peak emission moves to higher energies.  For example, the power emission per polarization is larger for a spin-$1/2$ particle (such as a neutrino) than for a spin-1 particle (such as a photon).  For black holes with masses $M_{\textrm{PBH}} \gtrsim 9.5\times 10^{13}\;{\rm kg}$, which are too large (and thus too cold) to emit ultrarelativistic $e^\pm$ pairs, most ($\sim 81$\%) of their energy emission is in neutrinos, which do not heat the IGM.  Black holes with $M_{\textrm{PBH}} \lesssim 9.5\times 10^{13}\;{\rm kg}$, in contrast, emit $\sim 45$\% of their energy in $e^\pm$ pairs, compared to only $\sim 9$\% in photons \cite{Page:1976df}.
Black holes with masses $M_{\textrm{PBH}} \lesssim 4.5 \times 10^{11}\;{\rm kg}$ will emit relativistic $\mu^\pm$ pairs, and black holes with progressively smaller masses will emit more and more massive particles.  At sufficiently small $M_{\textrm{PBH}}$ these will contribute to the photon, neutrino and electron emission by secondary decays and hadron jets.  By a coincidence between IGM physics and black hole physics, the black holes that are evaporating near the present epoch emit their photons into the window where the IGM optical depth is strongly dependent on redshift.

\section{IGM physics}\label{s:IGMphysics}

The mean IGM evolution is determined by tracking the kinetic temperature of the gas $T_K$, the ionization fraction $x_i$, and the spin temperature $T_S$.  
The evolution of $T_K$ and $x_i$ with redshift is given by differential equations, with initial conditions taken from standard cosmology just before recombination, and with additional terms to account for exotic energy injection mechanisms. The spin temperature $T_S$ is then determined algebraically. Neutral hydrogen is visible in 21cm whenever the spin temperature $T_S$ differs from the CMB temperature, $T_{\textrm{CMB}}$, and due to the contrast, the signal will show up in either emission or absorption.  The sky-averaged brightness temperature $T_b$ therefore carries information about the mean gas temperature and ionization state of the universe.
However, because of the many bright foregrounds expected for 21cm experiments (discussed in more detail in \S \ref{ss:foregrounds}), an absolute all-sky signal will be difficult to detect.  An easier target is a statistical detection of the power spectrum of the 21cm brightness temperature perturbations.  By comparing the power spectrum of the 21cm signal with and without a contribution from PBHs, we can retain the ability to discriminate among models within the limitations of realistic experiments.  We discuss the power spectrum calculation in \S \ref{s:ps}.

\subsection{Mean IGM evolution}

The homogeneous IGM is described by the mean kinetic temperature $T_K$, the spin temperature $T_S$, and the ionization fraction $x_i$.  The mean kinetic temperature of the gas evolves according to \cite{Furlanetto:2006wp,Padmanabhan:2005es}
\be\label{eq:dTkdt}
\frac{\ud T_K}{\ud t} = -2 H(z) T_K + \frac{x_i(z)}{\eta_1 t_\gamma}(T_{\textrm{CMB}} - T_K) + \frac{\chi_h \epsilon}{k_B}
\ee
where $H(z)$ is the Hubble parameter at redshift $z$, $\eta_1 = 1 + f_{He}+x_i$ with $f_{He}$ the helium fraction (defined by $n_{He}/[n_{H} + n_{He}]$) \footnote{The Recfast code \cite{Seager:1999bc} uses the alternative convention $f_{He} \equiv n_{He}/n{H}$; we correct for this discrepancy in our calculations.}, and 
\be
t_\gamma = \frac{3 m_e c}{8\sigma_T U_{\rm CMB}}
\ee
with $m_e$ the electron mass, $\sigma_T$ the Thomson cross section, and $U_{\rm CMB}$ the CMB energy density at redshift $z$.  
The first term on the right-hand side of Equation (\ref{eq:dTkdt}) accounts for the redshifting of kinetic energy with the expansion of the universe.  The second term includes the effect of IGM heating by scattering of CMB photons from hydrogen ions.
The final term in (\ref{eq:dTkdt}) takes account of heating by exotic energy injection into the IGM.  The parameter $\epsilon$ is an energy injection rate per baryon, and $\chi_h$ is the fraction of the energy that goes into heating the IGM.  We describe in more detail below how $\epsilon$ is determined for the PBH population.

The ionized fraction $x_i$ obeys \cite{Furlanetto:2006wp,Padmanabhan:2005es}
\be\label{eq:dxidt}
\frac{\ud x_i}{\ud t} = - \alpha (T_K) x_i^2 n_H + 
\frac{\chi_i \epsilon}{E_{ion}}
\ee
where the first term on the right-hand side includes hydrogen recombination through an effective coefficient $\alpha (T_K)$, which depends on the kinetic temperature of the gas.  Following \cite{Seager:1999bc} we use the Case B coefficient from Table I of
\cite{Pequignot:1991}.  The second term in Equation (\ref{eq:dxidt}) includes the ionizations produced by an exotic energy injection mechanism. As in Equation (\ref{eq:dTkdt}),  $\epsilon$ is the energy injection rate per baryon, $\chi_i$ gives the fraction of energy going into ionizations, and $E_{ion} = 13.6$ eV is the ionization threshold of hydrogen.

The spin temperature of the neutral hydrogen gas depends on competition among collisional excitation \cite{Purcell:1956}, heating by CMB photons, and interactions with Ly-$\alpha$ photons \cite{Wouthuysen:1952,Field:1958,Hirata:2005mz}.  The temperature itself is defined by the ratio between the occupation numbers $n_1$ and $n_0$ of the singlet and triplet hyperfine spin states, through
\be
\frac{n_1}{n_0} = 3 \exp \left( -\frac{T_S}{T_*} \right),
\ee
where $T_* = 0.068$ K is the energy splitting between hyperfine states in temperature units.
In equilibrium, the spin temperature $T_S$ is given by
\be\label{eq:TS}
\frac{1+x_c+x_\alpha}{T_S} = 
 \frac{1}{T_{\textrm{CMB}}} + \frac{x_c}{T_K} + \frac{x_\alpha}{T_c},
\ee
where the parameters $x_c$ and $x_\alpha$ are the collisional and Wouthuysen-Field \cite{Wouthuysen:1952,Field:1958} coupling parameters, respectively, and $T_c$ is the effective color temperature of the radiation field \cite{Hirata:2005mz,Chen:2004}.  We omit the Wouthuysen-Field coupling in this work, since we only expect it to be important once the first luminous sources turn on at comparatively low redshift, at which point their heating effects will easily swamp the contribution from PBHs.  The collisional coupling coefficient $x_c$ is 
\be
x_c = \frac{4 \kappa_{1-0}(T_K) n_H T_*}{3 A_{10} T_{\textrm{CMB}}},
\ee
where $A_{10} = 2.87 \times 10^{-15}$ s$^{-1}$ is the spontaneous decay rate of the hyperfine transition.  The parameter $\kappa_{1-0}$ describes spin excitation through atomic collisions; its value has been tabulated as a function of gas temperature \cite{Allison:1969,Zygelman:2005}.  For the range $1\; {\rm K} < T_K < 300 \;{\rm K}$, we use the values in column 4 of Table II in \cite{Zygelman:2005}. For $T_K > 300\;{\rm K}$, we use the fitting formulae suggested in \cite{Zygelman:2005} for the data presented in \cite{Allison:1969}.

Using Equations (\ref{eq:dTkdt}), (\ref{eq:dxidt}) and (\ref{eq:TS}) to evolve $T_K$, $x_i$ and $T_S$, we calculate the differential 21cm brightness relative to the background \cite{Madau:1997,Barkana:2006ep}:
\begin{align} \label{eq:Tb}
T_b &= \frac{T_S-T_{\textrm{CMB}}}{1+z}\left(1-e^{-\tau}\right) \\
&\simeq  28\;{\rm mK}\; \left( \frac{\Omega_b h}{0.033}\right) \left( \frac{\Omega_m}{0.27}\right)^{-1/2}  \times \notag \\
& \qquad\qquad \left(\frac{1+z}{10}\right)^{1/2} \left(\frac{T_S-T_{\textrm{CMB}}}{T_S}\right)
\end{align}
where we assume $\tau \ll 1$ in the second line. 

\subsection{21cm power spectrum} \label{s:ps}

Measurements of the power spectrum of fluctuations in the 21cm signal will make it possible to extract statistical information about the structure in neutral hydrogen even if detailed imaging or tomography remain beyond the reach of experiments.

As evident in Equations (\ref{eq:TS}) and (\ref{eq:Tb}) above, the brightness of the 21cm signal depends on the density, temperature and ionization state of the hydrogen gas.  To trace the shape of the power spectrum as a function of redshift, we must track not only the density perturbations in the baryonic matter, but also the fractional perturbations in the ionization state and the temperature of the hydrogen gas and the factors that influence the couplings among them.

The fractional perturbation to the brightness temperature is defined as
\be
\delta_{21}({\bf x}) \equiv [\delta T_b({\bf x}) - \bar{\delta T_b}] / \bar{\delta T_b}.
\ee
The Fourier transform $\delta_{21}({\bf k})$ can be written as a sum of each contribution weighted by expansion coefficients related to the various couplings \cite{Furlanetto:2006wp,Furlanetto:2006jb}:
\be
\delta_{21}({\bf k}) = (\beta + \mu^2)\delta + \beta_H \delta_H + \beta_\alpha \delta_\alpha + \beta_T \delta_T,
\ee
where the $\delta_i$ are fluctuations in overdensity ($\delta$), neutral fraction ($\delta_H$), Lyman-$\alpha$ coupling strength ($\delta_\alpha$) and temperature ($\delta_T$).  The expansion coefficients are
\begin{eqnarray}
\beta & = & 1 + \frac{x_c}{x_{tot} (1 + x_{tot})} \\
\beta_H & = & 1 + \frac{x_c^{HH} - x_c^{eH}}{x_{tot}(1 + x_{tot})} \\
\beta_\alpha & = & \frac{x_\alpha}{x_{tot} (1 + x_{tot})} \\
\beta_T & = & \frac{T_\gamma}{T_K - T_\gamma} + \frac{1}{x_{tot} (1 + x_{tot})} \times \notag \\
&\; & \left( x_c^{eH} \frac{\ud \ln{\kappa_{10}^{eH}}}{\ud \ln{T_K}} + x_c^{HH} \frac{\ud \ln{\kappa_{10}^{HH}}}{\ud \ln{T_K}} \right),
\end{eqnarray}
where the $x_c$ are collisional coupling coefficients \cite{Pritchard:2007}, for which $x_c = x_c^{eH} + x_c^{HH}$ and $x_{tot} = x_c + x_\alpha$. The parameters $\kappa^{eH}_{10}$ and $\kappa^{HH}_{10}$ are rate coefficients for spin de-excitation from electron/hydrogen and hydrogen/hydrogen collisions, respectively \cite{Furlanetto:2006jb}.

The power spectrum of 21cm fluctuations contains contributions proportional to the mean brightness temperature of the 21cm line over the whole sky as well as to the fluctuations in the spin temperature \cite{Lewis:2007}.  Taking a simplified estimate in which we consider only the growing-mode perturbation on large scales, we can write all perturbations as pure time-delay perturbations, which means each separate perturbation, and hence the 21cm power spectrum, is proportional to the matter power spectrum $P_{\delta \delta}$.  As we discuss in more detail below, this approximation leads to a small error when the brightness temperature is near zero, but provides sufficient accuracy for our purposes.  Then, the 21cm power spectrum can be written
\be
P_{21}(k,\mu) = \bar{\delta T_b^2} (\beta' + \mu^2) P_{\delta \delta}(k),
\label{eq:P21}
\ee
where $\mu = \cos (\theta)$ accounts for the angle between the wavevector ${\bf k}$ and the line of sight, and we have
\be
\beta' = \beta + \beta_T g_T - \frac{\beta_H \bar{x_i} g_i}{(1-\bar{x_i})} + \theta_u \beta_\alpha.
\ee
Here, $g_i(z) \equiv \delta_i/\delta$ and $g_T(z) \equiv \delta_T/\delta$ are defined for convenience to compare the fluctuations in the ionization and temperature to the matter overdensity, and $\theta_u$ encodes the uniformity of the energy deposition.  It is defined such that the energy deposition rate is proportional to $(1+\theta_u \delta)$ so that $\theta_u=0$ for uniform deposition.  Recall that we are neglecting the contribution of Wouthuysen-Field coupling in this work, so we are effectively imposing $\beta_\alpha=0$.

In practice, the power spectrum is calculated by evolving $g_i$ and $g_T$ with redshift along with the mean global properties of the IGM.  The evolution equations are \cite{Furlanetto:2006wp}
\begin{eqnarray}
\frac{\ud g_T}{\ud z} & = & \frac{g_T - 2/3}{1+z} + \frac{\bar{x_i}}{\eta_1 t_\gamma} \frac{g_T T_\gamma - g_i(T_\gamma - \bar{T_K})}{\bar{T_K} (1+z) H(z)} \notag \\
&\;&+ \frac{2}{3} \frac{\eta_2 \epsilon}{\eta_1 k_B \bar{T_K}} \chi_h \frac{(1-\theta_u) + g_T}{(1+z) H(z)} \label{eq:dgTdz} \\
\frac{\ud g_i}{\ud z} & = & \frac{g_i}{1+z} + \frac{\alpha \bar{x_i} \bar{n_H} (1 + g_i + \alpha' g_T))}{(1+z) H(z)} \notag \\ 
&\;& + \eta_2 \frac{\epsilon}{E_{ion}} \frac{\chi_i}{\bar{x_i}} \frac{(1 - \theta_u) + g_i}{(1+z) H(z)} \label{eq:dgidz},
\end{eqnarray}
where we have included bars over quantities that are global averages, for clarity.  The quantity $\alpha$ is the hydrogen recombination coefficient and $\alpha' \equiv \ud \ln \alpha / \ud \ln T_K$ \cite{Furlanetto:2006wp}.

\subsection{Energy injection and the photon spectrum}

The PBHs influence the IGM through the direct emission of photons and $e^\pm$ pairs.  Given the black hole emission power in these channels we must take into account the details of photon and $e^\pm$ absorption by the IGM in order to give the power absorption per baryon $\epsilon$ which appears in the evolution equations for $T_K$ and $x_i$.

\begin{figure}
\center{\includegraphics[width=3.5in,angle=0]{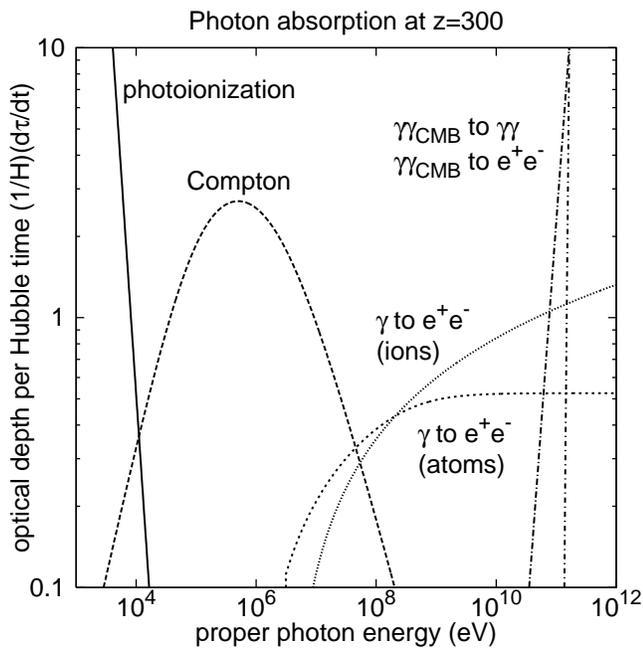}}
   \caption{Optical depths per time for various photon-IGM processes, in units of the Hubble time, at $z=300$, assuming a neutral IGM.  These include processes which deposit energy directly into the IGM (pair production and photoionization), processes which redistribute photons ($2\gamma\to 2\gamma$) and ones that do both (Compton).  At very low energies, photoionization is the dominant process; at very high energies, $e^\pm$ pair production dominates.}
\label{f:phot_igm}
\end{figure}

\begin{figure}
\center{\includegraphics[width=3.5in,angle=0]{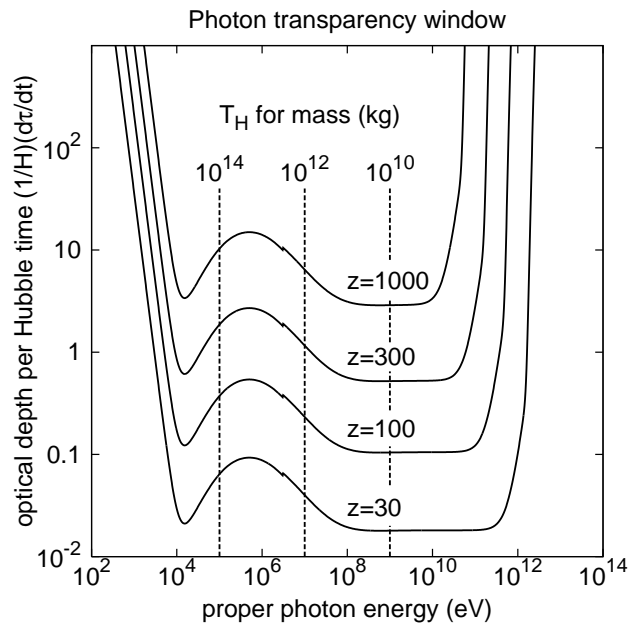}}
   \caption{Total optical depth for photon-IGM processes at various redshifts.  For photon energies where $(\ud\tau/\ud t)/H < 1$, the IGM is essentially transparent.  The Hawking temperature $T_H$ for various PBH masses is also shown.  Because of the graybody factors, the peak in $e^\pm$ energy emission is at energies slightly above $k_B T_H$, and the peak in the photon energy emission higher still.}
\label{f:window}
\end{figure}

There are a number of processes that allow photons to deposit energy in the IGM, all of which have frequency dependencies, which themselves change with redshift and with the ionized fraction.  We follow the tabulation of optical depths given in \cite{Zdziarski:1989} for the following processes: photoionization, Compton scattering, pair production from atoms and ions, photon-photon scattering, and single and double pair production from CMB photons.  
Individual rates for each process are shown at a fixed redshift in Figure \ref{f:phot_igm}, and the total photon-IGM interaction rate as a function of redshift is shown in Figure \ref{f:window}.
At all redshifts, below roughly 100 eV, the optical depth is dominated by photoionization, while at high energies it is dominated by pair production from CMB photons.  Between the two, the absorption of photons depends strongly on energy.  At lower redshifts, a ``transparency window'' for photon-IGM interactions emerges \cite{Chen:2003gz,Furlanetto:2006wp,Padmanabhan:2005es}. The ``bump'' visible in the plot is due to Compton scattering, and photons that fall into the trough at slightly higher energies are largely unabsorbed by the IGM.

To enable us to fully account for the IGM/photon physics, we follow the photon population in $\sim 10^3$ comoving energy bins between 1 and $10^{12}$ eV.  (We bin by comoving energy so that redshifting alone does not move photons from bin to bin.)  Careful accounting is necessary because of a numerical coincidence between the transparency window (the trough between $10^6-10^8$ eV at $z=300$) and the Hawking temperature of long-lived primordial black holes.  The black hole mass that is most interesting from the perspective of IGM physics is that corresponding to black holes that are just evaporating away at the present epoch, for these black holes deposit almost all of their initial mass-energy in the IGM.  A black hole that is just evaporating today had an initial mass of
$5 \times 10^{11}$ kg \cite{Page:1976df}, corresponding by Equation (\ref{eq:HawkT}) to a temperature of 20 MeV, which is nearly constant until the very end of the hole's life.  Referring to Figure \ref{f:window} reveals that these black holes are emitting a substantial fraction of their photons near the transparency window.
Therefore, most photons emitted by these black holes will not be absorbed for many Hubble times, if at all.  In this case, treating black hole emission as absorbed ``on the spot'' would be a poor approximation.  To properly treat black hole emission near this mass range, then, we follow the full photon population over time.

Different photon/IGM processes deposit energy in the IGM in different ways.  The pair production processes (from atoms or ions, and single and double pair production from CMB photons) all deposit their energy by converting it into $e^\pm$ pairs.  We count all of the energy going into the $e^\pm$ pair (including the rest masses of the $e^\pm$ themselves) as injection into the IGM via the $\epsilon$ parameter above.  When deposited in the IGM, the energy is partitioned between heating, ionization, and HI excitation.  The fractions of the initial electron energy that go into heating and ionization are denoted by $\chi_h$ and $\chi_i$, which were introduced in (\ref{eq:dTkdt}) and (\ref{eq:dxidt}).  These fractions have been computed in detail \cite{Shull:1985}; for our work we employ the approximations \cite{Chen:2003gz,Furlanetto:2006wp}
\be
\chi_i = \frac{1-x_i}{3},\quad \chi_h = \frac{1+2x_i}{3}.
\label{eq:chii}
\ee
We treat photoionization within the same framework.  The initial photoionization event is different than heating or ionization caused by the energetic electrons produced by pair production.  Nonetheless, except for photons near the photoionization threshold of 13.6 eV, the majority of the initial photon energy is given to the secondary photoelectron, which then heats and ionizes subsequent hydrogen atoms and ions in the same way as the electrons produced by pair production.  Tracking the full photon population allows us to properly treat photons whose first interaction occurs long after they are created, but for all of the pair production processes, we assume that the entire photon energy is lost on this first interaction. 

In contrast to the pair production processes, other processes largely redistribute photons within the photon population.  These photons do not deposit their energy in the IGM directly, though they may do so after downshifting to lower energies.  The main example of such a process is two-photon scattering
$\gamma\gamma_{\rm CMB}\to\gamma\gamma$ of energetic photons from the CMB.  
We assume that each interaction evenly divides the initial energy between the two photons, and so this produces a cascade of photons at successively lower energies.  These may eventually interact, or fall into the transparency window and be lost.  To a certain extent, Compton scattering also redistributes photons among the energy bins.  Unlike the other processes we consider, each Compton scatter may only reduce the energy of the photon by a small amount, and so we cannot make the approximation that all of the initial photon energy is deposited in the IGM.  We deposit an energy-dependent fraction (given in \cite{Zdziarski:1989}) of the photon's energy in the form of hot electrons, and decrement the energy of the photon itself.  Thus Compton scattering both heats and ionizes the IGM and modifies the photon population.

\section{Numerical results}\label{s:numerical}

We have developed a modular code that simultaneously: (a) evolves the PBH population; (b) calculates the energy output in photons and other particles;  (c) accounts for partial absorption and free streaming due to the transparency window (evolved with redshift); (d) partitions the energy injection into heating and ionization; (e) evolves the sky-averaged quantities; and (f) calculates the perturbations in heating and ionization for the 21cm power spectrum.

For initial conditions, we use a modified version of the RECFAST \cite{Seager:1999bc} code starting at redshift $z=10^4$, and then use that code to continue to follow the evolution of the mean temperature and ionization and their perturbations up to a redshift $z=300$. After that point, we use a simplified IGM evolution code to implement Equations (\ref{eq:dTkdt}), (\ref{eq:dxidt}), (\ref{eq:dgTdz}), and (\ref{eq:dgidz}).  Both RECFAST and the simpler IGM evolution code use an additional code module which tracks the photon population, and which computes the energy injection rate from photons and $e^\pm$ pairs which feeds back into the IGM evolution equations.

For the purpose of this calculation, we begin with a population of PBHs at a single mass and a given number density.  The possible spectra of masses of PBHs is discussed in \S \ref{s:formation}.

We follow the evolution of the population of PBHs as they lose mass to Hawking radiation, injecting their energy into the IGM.  Based on the PBHs' effect on the 21cm power spectrum in a range of redshifts and the expected observational limitations of future experiments, we are able to obtain a limit on the density of PBHs at each initial mass in our range of interest.

For most of the mass range of interest for 21cm experiments, the PBHs have evaporated completely before the present day.  As can be seen in Equation (\ref{eq:HawkT}), the Hawking temperature diverges as the PBH mass goes to zero.  There has been much discussion in the literature as to the phenomenology of the final stages of Hawking evaporation, with some suggesting that the PBHs may leave a Planck-mass relic \cite{Maeda:1986,Bowick:1988xh,Coleman:1991}, and others proposing a connection between the late stages of PBH evaporation and signatures of extra dimensions \cite{Kavic:2008qb}.  For our purposes, the details of the final moments are not important, since only a very small fraction of the PBH's total energy is emitted during that time.  We ``turn off'' the Hawking evaporation when the PBHs are within one year of complete disappearance.

We find there are several regimes of PBH mass ranges that have distinct effects on the evolution of the IGM:
\begin{itemize}
\item {\bf \boldmath$M_{\textrm{PBH}} \lesssim 5 \times 10^{10}$ kg:}  In this regime, the evaporation occurs early enough to alter the late stages of recombination, but these PBHs are not well suited to constraints with Dark Age 21cm observations. 
\item {\bf \boldmath$5 \times 10^{10}$ kg $\lesssim M_{\textrm{PBH}} \lesssim 10^{11}$ kg:}  PBHs in this range of masses evaporate in a range of redshifts ($30 \lesssim z \lesssim 90$) during which the fiducial sky-averaged 21cm brightness temperature signal is visible in absorption against the CMB.  At sufficient number densities, these PBHs would raise the brightness temperature in 21cm, effectively decreasing the strength of the 21cm signal during this time.  An anomalously small (in absolute value) 21cm brightness temperature on the sky would indicate extra energy injection in this regime.
\item {\bf \boldmath$M_{\textrm{PBH}} \sim 10^{11}$ kg:}  The evaporation of PBHs with initial masses near $10^{11}$ kg have the most dramatic effect on the expected 21cm signal.  The peak in their energy injection would occur at sufficiently low redshifts that, at high densities, they could raise the spin temperature above the temperature of the CMB, making the 21cm signal appear in emission rather than in absorption.  For the smallest masses, at number densities not yet ruled out by current data, the emission signal can be several mK.  If the all-sky signal were to switch to emission before the effects of star formation were expected to significantly alter the signal (stars can bring the 21cm signal into emission for redshifts less than $z \sim 15$ \cite{Furlanetto:2006tf}), this would be a strong indication of exotic physics in the Dark Ages~\cite{Furlanetto:2006wp}.
\item {\bf \boldmath$10^{11}$ kg $\lesssim M_{\textrm{PBH}} \lesssim 10^{14}$ kg:}  PBHs created in this mass range would still exist today, as their evaporation timescale is longer than the current age of the universe.  For these PBHs, the Hawking radiation can be well approximated by a constant rate of energy injection; their effect on the IGM would be very similar to that of decaying dark matter (see \cite{Furlanetto:2006wp} for a discussion of the effects of such a scenario).  Since the power and Hawking temperature are both small for such massive PBHs, their effect on the IGM would be less pronounced than that of their less massive counterparts.  They would also lack the sudden, short-lived increase in signal that can be seen for PBHs that complete their evaporation in the Dark Ages.
\item {\bf \boldmath$M_{\textrm{PBH}} \gtrsim 10^{14}$ kg:}  At very high masses, the Hawking temperature is too low to allow for the production of $e^\pm$ pairs.  These black holes emit cooler photons which do not efficiently inject energy into the IGM.  Therefore a greater number density of these black holes can be tolerated without changing the mean IGM evolution significantly.

\end{itemize}

For black holes with masses above $10^{14}$ kg, which do not emit $e^{\pm}$ pairs, the energy injection rate is roughly constant.  This allows for a direct comparison with decaying dark matter models in which the dark matter lifetime is much longer than the age of the universe, which give a roughly constant energy injection rate per baryon.  Therefore we can find a dark matter decay model corresponding to a given large-mass PBH evaporation model.  For cases in which all the dark matter is in the form of decaying dark matter particles, the correspondence is:
\begin{align}
& \left( \frac{M_{\textrm{PBH}}}{10^{15} \textrm{ kg}} \right)^{-2} \left( \frac{n_{\textrm{PBH}}}{1.2 \times 10^{-42} \textrm{ m}^{-3}} \right) \notag \\
& \qquad = \left( \frac{\Gamma_X}{3.18 \times 10^{-30} \textrm{ s}^{-1}} \right) 
\end{align}
for $M_{\textrm{PBH}} > 10^{14}$ \textrm{ kg},
where $\Gamma_X$ is the decay rate of the dark matter particle.  As a corollary, observational consequences in 21cm of PBH populations are a function of $M_{\rm PBH}^{-3} \Omega_{\rm PBH}$  for $M_{\rm PBH} > 10^{14}$ kg.  For smaller PBHs, the emission of $e^{\pm}$ pairs dominates the energy injection, and this scaling no longer applies.

To determine the detectability of PBHs evaporating in the Dark Ages, we simulate their effects in three potentially observable quantities.  The first is the overall ionization state of the IGM, the second is the high-redshift 21cm all-sky brightness temperature, and the third (and most relevant for near-term experiments) is the 21cm power spectrum.  We discuss each in turn in the following subsections.

To illustrate the effects of PBHs in different mass regimes, we will use three example PBH models, whose effects on the IGM are plotted in Figures \ref{f:ion}, \ref{f:bright} and \ref{f:allps}.  The parameters of the three cases are described in Table \ref{t:models}.

\begin{table}[h!b!p!]
\caption{Example primordial black hole models for Figures \ref{f:ion}, \ref{f:bright} and \ref{f:allps}.}
\begin{tabular}{c | c | c | c}
\hline\hline
Model & $M_{\textrm{PBH}}$ (kg) & $\Omega_{\textrm{PBH}}$ & Plot key \\
\hline
A & $10^{13}$ & $1.3 \times 10^{-5}$ & green, dot-dashed\\
B & $10^{11}$ & $8.2 \times 10^{-12}$ & blue, dashed \\
C & $5 \times 10^{10}$ & $2.7 \times 10^{-12}$ & magenta, dotted \\
\hline
\end{tabular}
\label{t:models}
\end{table}

\subsection{Mean ionized fraction}

\begin{figure}
\center{\includegraphics[height=3.3in,angle=90]{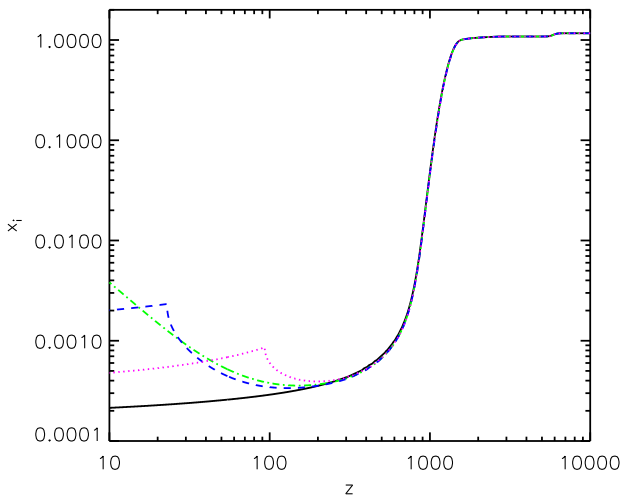}}
   \caption{The mean ionized fraction vs. redshift for the fiducial case of no PBH energy injection (thick solid line) and three PBH cases, described in Table \ref{t:models}: $M_{\textrm{PBH}}=10^{13}$ kg at $\Omega_{\textrm{PBH}}=1.3 \times 10^{-5}$ (Model A; green, dot-dashed), $M_{\textrm{PBH}}=10^{11}$ kg at $\Omega_{\textrm{PBH}}=8.2 \times 10^{-12}$ (Model B; blue, dotted) and $M_{\textrm{PBH}}=5 \times 10^{10}$ kg at $\Omega_{\textrm{PBH}}=2.7 \times 10^{-12}$ (Model C; magenta, dotted).  These models correspond to marginally detectable deviations in the power spectrum, as discussed in Equation (\ref{eq:pserror}) in \S \ref{ss:powerspectrum}.}
\label{f:ion}
\end{figure}

As the PBHs evaporate, the energy they inject into the IGM will increase the hydrogen ionized fraction.  To examine the importance of this effect, we simulated the ionization history of the IGM, following Equation (\ref{eq:dxidt}) with the energy deposition for ionization determined by Equation (\ref{eq:chii}).  For the models we considered, which would affect the 21cm power spectrum at potentially detectable levels (see \S \ref{ss:powerspectrum}), the ionization fraction did not increase to a level sufficient to reionize the universe; generally, ionization fractions of the order of $10^{-3}$ are only reached at redshifts low enough that the formation of stars and galaxies would likely start to become important.  We plot some example ionization histories in Figure \ref{f:ion}.

The models we plot here are chosen to reflect the range of phenomenology that can be expected from PBH evaporation in the regimes in which it would strongly affect the IGM evolution.  In addition to the no-PBH case, we plot two cases in which the PBH evaporation completes during the Dark Ages, giving the strongest signal as the temperature and power of the Hawking radiation increase greatly in the final stages of evaporation, and one case in which the evaporation is not complete by the present day.  The parameters of these models are summarized in Table \ref{t:models}. In Figure \ref{f:ion}, the peaks in each of the two lowest-mass models occur at the redshift at which the evaporation finishes in each model.

One may ask if the alteration in the ionization history would be detectable as a change in the integrated Thomson scattering optical depth measurement ($\tau_{\textrm{LSS}}$) in CMB experiments.  For the models we focused on, with potentially detectable differences in the 21cm power spectrum (as determined by equation [\ref{eq:pserror}]), we find that the change in $\tau_{\textrm{LSS}}$ is at best on the order of $10^{-3}$, whereas the contribution to $\tau_{\textrm{LSS}}$ from the reionization of the universe is two orders of magnitude higher.  Current experiments cannot reach this precision, and uncertainties in the process of cosmological reionization would likely prevent exotic physics from being constrained in this way in the future: for example, a change in the reionization redshift from 6.1 to 6.0 would result in a change in $\tau_{\textrm{LSS}}$ of order $10^{-3}$, and non-instantaneous reionization scenarios complicate the picture further.  The fact that the optical depth measurement is an integrated signal over the history of the universe since recombination gives it a disadvantage over future 21cm experiments, where measurements will be possible at a large number of redshifts.  With CMB polarization measurements from the Planck satellite, we expect to be able to constrain energy injection rates of $ \sim 10^{-15}$ eV/s/baryon at $z=1000$ \cite{Padmanabhan:2005es}, whereas with future 21cm power spectrum measurements we can expect to do around three orders of magnitude better \cite{Furlanetto:2006wp}, and to measure at several redshifts.

Since the 21cm signal depends on the ionization state and the temperature of the gas, the features in the ionization history are echoed in the 21cm brightness temperature signal, which we discuss next.

\subsection{Sky-averaged 21cm brightness temperature}

The ultimate realization of the power of 21cm observations would be in the form of 21cm tomography of the high-redshift intergalactic medium.  A radio telescope observing the full sky at $21 \textrm{cm} (1 + z)$ could, in principle, have a complete map of the neutral hydrogen at that redshift, and could obtain a 3D image of structure formation by combining slices at each $z$.  While there are significant observational challenges involved in true tomography (some of which we discuss in \S \ref{s:observation}), even a sky-averaged mean signal of the 21cm brightness temperature, which would contain information about the overall state of the neutral hydrogen at a range of redshifts, would significantly improve upon current probes of the IGM at high redshifts.

\begin{figure}
\center{\includegraphics[height=3.3in,angle=90]{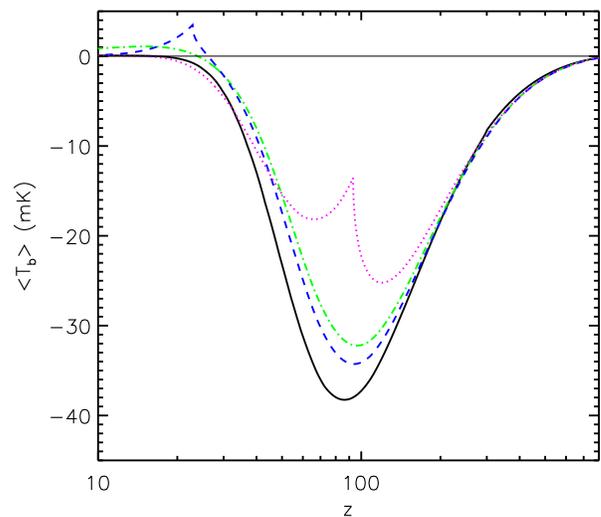}}
   \caption{The sky-averaged brightness temperature vs. redshift for the fiducial case of no PBH energy injection (thick solid line) and three PBH cases from Table \ref{t:models}, as in Figure \ref{f:ion} above.  The thin horizontal line is at a 21cm brightness temperature of zero (no signal).}
\label{f:bright}
\end{figure}

Following Equation (\ref{eq:Tb}), we track the mean 21cm brightness temperature as a function of redshift in our simulations.  Although there are still many observational difficulties for this measurement (including major foregrounds and challenges to instrument design, discussed in more detail in \S \ref{s:observation}), this result gives an impression of the effects of PBH energy injection on the 21cm observables, and experiments currently under way may be sensitive to strong gradients in the sky-averaged signal.  In Figure \ref{f:bright}, we plot the brightness temperature induced by the no-PBH case as well as by the three example models plotted previously in Figure \ref{f:ion}.

The two low-mass cases, in which the final evaporation occurs in the Dark Ages, show conspicuous spikes in the brightness temperature at the redshifts of evaporation.  For the lowest-mass case, $M_{\textrm{PBH}} = 5 \times 10^{10}$ kg (Model A), the effect is to decrease the 21cm signal in absolute value by tens of mK; this discrepancy should be noticeable if future experiments can measure the mean signal to sufficient accuracy.  The signal from the intermediate case (Model B) would potentially be much more striking, as the signal would switch from absorption to strong emission at a redshift above that at which ordinary star-formation is expected to bring the spin temperature above the CMB temperature.  The highest-mass case (Model C), in which evaporation has not occurred by  the present day, would also significantly alter the mean signal over a range of redshifts, though there would not be as dramatic a discrepancy.

It must be noted that the ``spikes'' in the ionization and brightness temperature signals depend upon the simultaneous evaporation of all PBHs --- this kind of signal would only occur if the PBHs were formed with a very narrow mass spectrum.  For a broader range of masses, the peaks would be smoothed out and the signals less striking.  Therefore the absence of a spike in the sky temperature would not necessarily disfavor the existence of PBHs with a broad mass distribution.

\subsection{21cm power spectrum} \label{ss:powerspectrum}

\begin{figure}
\center{\includegraphics[height=3.3in,angle=90]{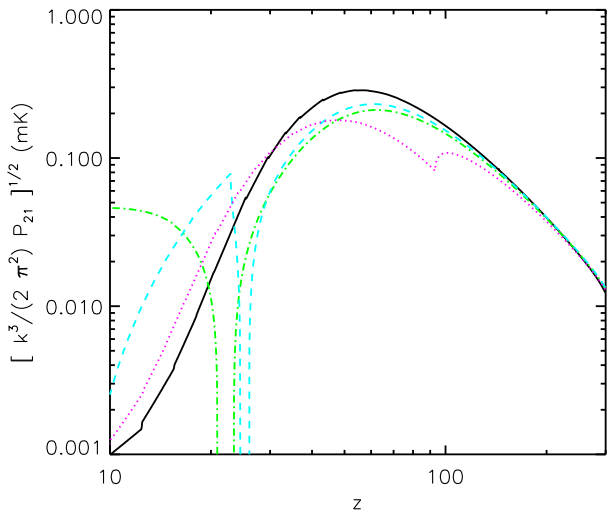}}
   \caption{The 21cm power spectrum for the fiducial case of no PBH energy injection (thick solid line) and the three PBH cases from Table \ref{t:models}, as in Figures \ref{f:ion} and \ref{f:bright} above.}
\label{f:allps}
\end{figure}

Given the observational difficulties inherent in 21cm tomographic imaging of early structures and in all-sky brightness temperature measurements, some of the near-term experiments are instead aimed at detecting the 21cm fluctuations statistically by measuring the 21cm power spectrum.  Realistic near-term experiments should have the capability to use the 21cm power spectrum to study the process of reionization \cite{Carilli:2005nj} and estimate cosmological parameters \cite{Mao:2008ug}, and we find that PBH evaporation can be strongly constrained by more ambitious future radio interferometers.

The ability to distinguish models including exotic physics from the standard IGM evolution depends upon how precisely the power spectrum can be determined.  We use an estimate of the error in the power spectrum measurement for a hypothetical future experiment to test the observability of the alteration in the IGM evolution due to our PBH models.  The error depends on several different parameters of the experiment in question.  An ideal radio interferometer would have a long maximum baseline ($R_{\textrm{max}}$), a high effective area ($A_e = N \times A$, where $N$ is the number of receivers and $A$ is the area of each), and a high covering factor ($f_{\textrm{cov}} \equiv A_e / (\pi R^2_{\textrm{max}})$).  A large bandwidth ($B$) and a long integration time ($t_{\textrm{int}}$) also decrease the error.  The frequency of the observation comes into the equation in two ways, and in both cases, higher redshifts give higher errors.  The error is proportional to $(1+z)$, but also to the sky temperature $T_{\textrm{sky}}$, which is a function of frequency.  The foreground is dominated by Galactic synchrotron, which roughly follows $T_{\textrm{synch}} \sim (\nu/200 \textrm{ MHz})^{-2.8}$ K for the frequencies of interest.

For the purpose of placing a constraint, we consider an experiment with parameters similar to estimates for the proposed Square Kilometer Array (SKA), but operating at a much lower frequency in order to probe the higher-redshift universe \cite{Schilizzi:2007}.  We compare the parameters of the considered low-frequency experiment with those of SKA in Table \ref{t:obs_params}.  We assume an integration time of 1000 hr for each, and observations at wavenumber $k = 0.04$ Mpc$^{-1}$.

\begin{table}[h!b!p!]
\caption{Experimental parameters for proposed future low-frequency observation compared with expected SKA parameters.}
\begin{tabular}{| c | c | c |}
\hline
  & SKA & Future experiment \\
\hline\hline
 $R_{\textrm{max}}$ & 5 km & 5 km \\
\hline
 $f_{\textrm{cov}}$ & 0.01 & 0.25 \\
\hline
 bandwidth & 2 GHz\footnote{The bandwidth used for an individual observation is frequency-dependent; see \cite{Schilizzi:2007} for details.} & 50 MHz \\
\hline
 min. frequency & 100 MHz & 29 MHz \\
\hline
 max. $1+z$ & 14 & 50 \\
\hline
 $T_{\textrm{sky}}$ & 660 K & $10^4$ K \\
\hline
\end{tabular}
\label{t:obs_params}
\end{table}

From \cite{Furlanetto:2006wp}, the error at a wavenumber $k$ is given by:
\begin{eqnarray}
\sqrt{\frac{k^3 \delta P_{21}}{2 \pi^2}} & \sim & \frac{0.1 \textrm{ mK}}{\epsilon^{1/4}} \left(\frac{0.25}{f_{\textrm{cov}}} \right) \left( \frac{5 \textrm{ km}}{R_{\textrm{max}}} \right) \notag \\
&  \times & \left(\frac{k}{0.04 \textrm{ Mpc}^{-1}}\right)^{3/4} 
\left( \frac{T_{\textrm{sky}}}{10^4 \textrm{ K}} \right)  \left(\frac{50 \textrm{ MHz}}{B} \right)^{1/4} \notag \\
& \times & \left(\frac{1000 \textrm{ hr}}{t_{\textrm{int}}} \right)^{1/2} \left( \frac{1+z}{50} \right).
\label{eq:pserror}
\end{eqnarray}
The data are assumed to be binned in segments of logarithmic length $\epsilon k$.

In Figure \ref{f:allps}, we plot example power spectra from our PBH models alongside the fiducial no-PBH model to illustrate how the power spectra change with PBH energy injection.  The zeros in the power spectra plots correspond to the 21cm signal changing from absorption to emission, since, from Equation (\ref{eq:P21}), for our approximation, the power spectrum is proportional to the square of the 21cm brightness temperature.

We have made some approximations in order to compute these power spectra, and here we can comment on their accuracy.  Essentially we have taken the 21cm power spectrum to be proportional to the mean 21cm brightness temperature $T_b$.  However there is also a contribution from fluctuations in the spin temperature $T_S$.  Normally these are small but can become important when the 21cm mean signal switches from absorption to emission    and $T_b$ passes through zero.  We can use the more accurate calculation of \cite{Lewis:2007} to estimate our error.  The fractional error, relative to the no-PBH 21cm power spectrum, is approximately
\be
\frac{T_s}{T_s-T_\gamma} \frac{\Delta_{T_s}}{\Delta_{HI}}
\ee
using the notation of \cite{Lewis:2007}.  This is the ratio between the fluctuations in the spin temperature (which we have neglected) to those proportional to the mean brightness temperature (which we have included).  Using the results of \cite{Lewis:2007} for the standard cosmology (no PBHs) gives a fractional error of $\sim 1/7$.  While not suitable for percent-level precision observations, our approximations can be trusted to indicate when PBHs lead to a substantial deviation from the expected 21cm signal, and give reliable results at the $\sim 10$\% level.

Applying the criterion of Equation (\ref{eq:pserror}) to these power spectra, we can estimate the ability of future 21cm observations to place limits on populations of PBHs.  We discuss our results in the Section \ref{s:results}.

\subsection{Relic photon population}

\begin{figure}
\center{\includegraphics[height=3.3in,angle=90]{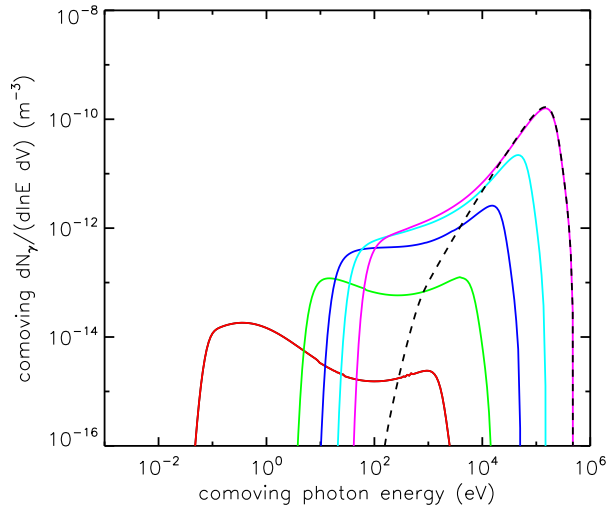}}
   \caption{Evolution of the photon population for Model A of Table \ref{t:models}, with $M_{\textrm{PBH}} = 10^{13}$ kg.  The solid curves, from left to right, are $z=4000$, $1000$, $300$, $100$ and $30$.  The dashed curve is the photon population at $z=30$ with all IGM-photon interactions turned off, given for comparison.}
\label{f:photons13}
\end{figure}

\begin{figure}
\center{\includegraphics[height=3.3in,angle=90]{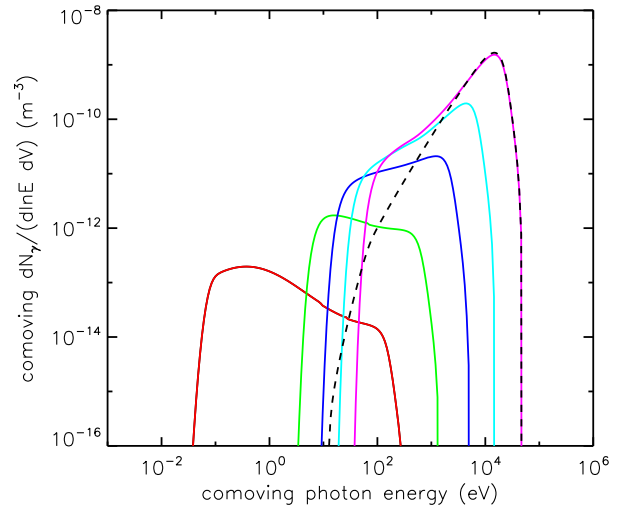}}
   \caption{Evolution of the photon population for a model with $M_{\textrm{PBH}} = 10^{14}$ kg and $\Omega_{\textrm{PBH}} = 1.3 \times 10^{-2}$.  The solid curves, from left to right, are $z=4000$, $1000$, $300$, $100$ and $30$.  The dashed curve is the photon population with all IGM-photon interactions turned off, given for comparison.}
\label{f:photons14}
\end{figure}

As a by-product of the IGM computations, we obtain predictions for the relic photon population produced by the PBHs.  These photons have an energy spectrum which is the Hawking spectrum, modified by the graybody factors, convolved over redshift, and then processed by IGM-photon interactions.

Plots of the photon population as a function of redshift are given in Figures \ref{f:photons13} and \ref{f:photons14}.  The populations in these two models follow a similar evolution pattern.  At very high redshift ($z > 1000$) the black holes emit photons near the Compton peak, which are subsequently downscattered to lower energies.  After recombination, the low-energy photons rapidly deposit all their energy in the IGM through photoionization of neutral hydrogen and disappear from the spectrum.  As the redshift decreases further, most photons are emitted into the transparency window.  The late-time photon population therefore relaxes to the unabsorbed Hawking spectrum, with a tail at lower energies from Compton downscattering.

The models shown in the Figures are chosen to be on the constraint surface for 21cm observations.  For these models, the relic photon population would be swamped by conventional astrophysical processes in the relevant energy range.  Hence 21cm observations could detect the presence of these PBH populations that would be unobservable from diffuse photon background measurements alone.

\section{Constraints on PBH population} \label{s:results}

The power spectrum of 21cm brightness temperature fluctuations will be a powerful probe of exotic physics in the Dark Ages, before the IGM is significantly heated and ionized by the first stars and galaxies.  We apply an estimate of the error in the power spectrum, from Equation (\ref{eq:pserror}), to our simulated power spectra for models of PBHs injecting energy into the IGM through Hawking radiation.  We find that future experiments will be able to place very tight constraints on PBHs that evaporate during the Dark Ages, improving on current limits by several orders of magnitudes in certain regimes, and demonstrating the feasibility of using 21cm observations to constrain exotic physics.

\begin{figure}
\center{\includegraphics[height=3.3in,angle=90]{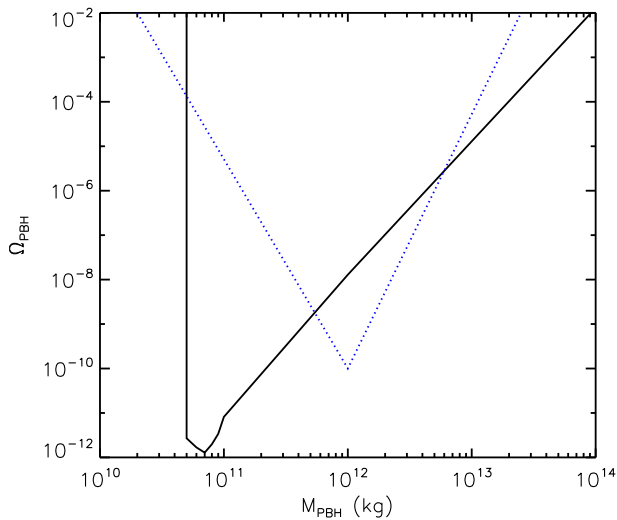}}
   \caption{The solid line shows our expected constraints on PBH mass and density parameter from a 21cm observational program with parameters described in Equation (\ref{eq:pserror}).  The blue dotted line shows the diffuse $\gamma$-ray background constraint from \cite{Carr:1994ar} for comparison.  PBH models with masses less than $\sim 5 \times 10^{10}$ kg are difficult to constrain with the 21cm observations we are proposing because they evaporate before the Dark Ages, and may interfere with the recombination process.}
\label{f:constraint}
\end{figure}

In Figure \ref{f:constraint}, we plot the constraint based on the simulated power spectra, using Equation (\ref{eq:pserror}) as an estimate of distinguishability, along with the PBH constraint from the diffuse gamma ray background \cite{Carr:1994ar}.

For the mass range in which the evaporation is not complete before the present day, the energy injection from evaporation remains roughly constant during the Dark Ages, and the constraint on the density parameter ($\Omega_{\textrm{PBH}}$) in PBHs is proportional to $M_{\textrm{PBH}}^3$.  For the case of a constant energy injection rate, the fact that the Hawking radiation power is proportional to $M_{\textrm{PBH}}^{-2}$ implies that the combination $M_{\textrm{PBH}}^{-3} \Omega_{\textrm{PBH}}$ determines the energy injection rate, so the scaling in that regime is as expected.

For intermediate masses, where the evaporation finishes during the Dark Ages, somewhat tighter constraints are possible due to the sharp increase in the power and Hawking temperature during the final stages of evaporation.

At the smallest masses, the evaporation occurs before recombination is complete; this case is not well constrained with Dark Age probes.

\section{Observational prospects} \label{s:observation}

The prospect of using 21cm observations to detect the transition from a neutral to ionized IGM has motivated a surge of activity in the observational sector, resulting in several ambitious low-frequency radio interferometry projects, some of which are already taking data.  

A review of the progress of current and planned observational projects can be found in \cite{Carilli:2005nj}.  For our purpose in discussing a possible future constraint on exotic physics in the Dark Ages, we imagine an ambitious future radio interferometry experiment that will measure the 21cm power spectrum at 0.1 mK precision at a redshift of 50.  The parameters of such an experiment (see Table \ref{t:obs_params}) are chosen to demonstrate the potential of 21cm observations rather than to point to a specific proposed experiment; our imagined experiment would be somewhat more advanced than the proposed SKA.

Sensitive study of neutral hydrogen in the Dark Ages with 21cm observations will require overcoming challenges both in instrumentation and in astrophysical foreground removal.  In terms of the instrumentation, it will be important to have tight control over the polarization response of the instrument and to carefully deal with frequency-dependent side-lobes \cite{Carilli:2005nj}.

An even bigger challenge to this observation will be the strong astrophysical and terrestrial foregrounds.  We discuss these briefly in the next section; for more details, see, e.g., \cite{Furlanetto:2006jb,Carilli:2005nj,Oh:2003jy,DiMatteo:2001gg,Shaver:1999gb,Wang:2005zj}.

\subsection{Foregrounds} \label{ss:foregrounds}

The radio sky at the frequencies of high-redshift 21cm observations is very bright; at around 200 MHz (corresponding to 21cm observations at $z \sim 6$), the temperature of the quietest parts of the sky can be hundreds of Kelvins, compared to an expected 21cm fluctuation signal on the order of tens of mK \cite{Furlanetto:2006jb}.  Radio synchrotron emission from our Galaxy is the dominant contribution to this foreground;  other contributions are free-free emission from our Galaxy and extragalactic contributions.  The extragalactic component could be particularly difficult to correct for in spatial measurements because the angular fluctuations are unknown and will swamp the 21cm signal \cite{Oh:2003jy}.

Frequency information obtained along individual lines of sight may alleviate the foreground problem to some extent.  Synchrotron and free-free emission should have smooth powerlaw spectra in frequency space, whereas the high-redshift 21cm signal from neutral hydrogen structures is expected to have more small-scale structure (see, e.g., \cite{Wang:2005zj}).  After point sources are excised, a smooth spectrum can be fit and subtracted, leaving only the 21cm fluctuations.  This would generally be done on a pixel-to-pixel basis, using a fitting function for the smooth foreground spectrum.  It would be especially effective in cases, such as with the SKA, where individual HII regions may be imaged.  These fully-ionized regions will have no signal in 21cm, and so will allow for a direct probe of the foregrounds \cite{Furlanetto:2006jb}.

Fluctuations in the ionosphere can also be problematic for low-frequency measurements, 
distorting images through refraction.  These distortions will likely be correctable except during periods of unusually high ionospheric activity \cite{Carilli:2005nj}.  Below frequencies of a few MHz, one approaches the plasma frequency of the ionosphere, and it becomes opaque.  This effectively prevents ground-based 21cm observations at redshifts higher than $z \sim 70$.

One of the most difficult foregrounds to correct for will be terrestrial radio interference, though there is much ongoing computational work being devoted to this problem.  Radio, television, and mobile phones all operate in the desired frequency range, as it is not a protected region of the spectrum.  For this reason, many of the current and upcoming observational efforts are occurring in remote locations with low population density.  To completely defeat the foregrounds from both the ionosphere and radio interference, some have suggested constructing observatories on the far side of the moon, with the lunar mass acting as a radio shield \cite{Carilli:2007eb,Lazio:2007zp}.

Tight instrumental control and careful foreground removal will be the biggest barriers to the use of 21cm observations for cosmology or early universe physics, and the ultimate goal of detailed imaging of early structure formation in neutral hydrogen may be thwarted by these challenges for the foreseeable future.  However, the foreground cleaning techniques that are currently being developed will make statistical measurement of the 21cm fluctuations through the power spectrum achievable with upcoming telescopes, and several projects are in development with this goal in mind, including the 21 Centimeter Array (21CMA; formerly PAST), the Murchison Widefield Array (MWA), and the Low Frequency Array (LOFAR) \cite{Carilli:2005nj}.

Complementary to power spectrum measurements will be the ongoing observational efforts to track the sky-averaged 21cm brightness temperature as a function of frequency (redshift).  While foregrounds will make an absolute measurement of the brightness temperature difficult, strong gradients over small frequency ranges will be detectable with the first generation of experiments \cite{Bowman:2007su}.  The hope is to use this technique to discover the transition between a neutral and an ionized IGM at the epoch of reionization, but if PBH evaporation occurs suddenly enough, the sharp increase in brightness temperature may also be a detectable signal.

The EDGES experiment, currently under construction, will use a single antenna at a radio-quiet site to measure the sky-averaged brightness temperature and search for gradients of around 1 mK/MHz, with a frequency resolution of between 0.1 and 1 MHz, in a range corresponding to the 21cm signal at $z<30$ \cite{Bowman:2007su,Bowman:pc}.  In our example PBH scenarios, the brightness-temperature peak from sudden evaporation would be somewhat below this detectability threshold, but future experiments along these lines may have the power to strongly constrain sharp PBH mass functions through the non-detection of peaks in the 21cm brightness temperature.

\section{Discussion}\label{s:discussion}

We have shown that future observations of the power spectrum of high redshift neutral hydrogen 21cm fluctuations have the potential to place strong constraints on the injection of energy into the intergalactic medium in the Dark Ages due to exotic physics, specifically in the case of evaporating primordial black holes.  The contribution of PBHs to the mass budget of the universe can potentially be tightly constrained in a range of masses from $5 \times 10^{10}$ kg to $10^{14}$ kg,  corresponding to PBHs whose final evaporation would occur during the Dark Ages when the universe is dominated by uncollapsed neutral hydrogen gas.  We estimate being able to constrain $\Omega_{\textrm{PBH}} \lesssim 10^{-12}$ at $M_{\textrm{PBH}} \sim 5 \times 10^{10}$ kg, and to improve upon current constraints derived from the $\gamma$-ray background for similar masses.

In order to estimate the effect of PBH evaporation on the IGM, we developed an extensive computer code for calculating the evolution and Hawking radiation of the PBHs, the transparency and absorption properties of the IGM, and the evolution of the 21cm signal both in terms of the sky-averaged brightness temperature and the fluctuation power spectrum.

Although the use of 21cm observations for cosmology in the Dark Ages will be technically challenging, the potential payoff for a wide range of astrophysical and cosmological applications easily justifies the current surge of interest in the field.

\section{Acknowledgments}
KJM acknowledges the support of the NSF GRFP.  The authors would like to thank Judd D. Bowman, Bernard Carr, Steve Furlanetto, Martin Haehnelt, Kazunori Kohri, Antony Lewis, Peng Oh and Jonathan Pritchard for helpful discussions.

\appendix
\section{Conventions}

In this section, we describe some notational conventions and definitions used in this work, and how they compare to related works cited herein.

The main source of differences among recombination calculations is the definition of the helium fraction, $f_{He}$.  The RECFAST code \cite{Seager:1999bc}, which we use for the highest redshifts in our calculation, defines $f_{He}$ as the helium-to-hydrogen number ratio:
\be
f_{He}^{\textrm{Recfast}} \equiv \frac{n_{He}}{n_H} = \frac{Y_p}{4(1-Y_p)} \label{eq:fHeRecfast}
\ee
where $Y_p$ is the primordial helium abundance (mass ratio), defined by $\rho_{He}/(\rho_H+\rho_{He})$, which we take to be 0.25.  Therefore, $f_{He}^{\textrm{Recfast}} = 1/12$.  An alternative definition of the helium fraction is the ratio between the number density of helium and the total number density of nuclei.  This definition is used by \cite[hereafter FOP]{Furlanetto:2006wp}, from which we derive late-time IGM evolution equations.  The numerical difference between the helium fraction defined this way and that of Recfast is small but potentially significant:
\be
f_{He}^{\textrm{FOP}} \equiv \frac{n_{He}}{n_{H} + n_{He}} = 1/13. \label{eq:fHeFOP}
\ee

Another possible source of confusion is the definition of the energy injection rate.  A simple option is to define it as the amount of energy injected per cubic centimeter, but it is often convenient to define it instead in terms of the amount of energy injected per baryon or per hydrogen nucleus.  Investigating the energy injection due to dark matter annihilation in the Dark Ages, the authors of \cite[hereafter PF]{Padmanabhan:2005es} define an energy injection rate 
\be
\epsilon^{\textrm{PF}} \equiv Q / n_H^0 \label{eq:epsPF}
\ee
 where $Q$ is the comoving energy injection rate per cubic centimeter, and $n_H^0$ is the comoving number density of hydrogen nuclei.  In a similar calculation, the authors of \cite{Furlanetto:2006wp} implicitly define the energy injection rate due to dark matter annihilation or decay as
\be
\epsilon^{\textrm{FOP}} \equiv Q / n_b^0 \label{eq:epsFOP}
\ee
where $n_b^0$ is the comoving number density of baryons.  Therefore, to compare the two calculations, one must keep in mind the conversion factor between the two energy injection rates:
\be
\epsilon^{\textrm{PF}} = (1+4 f_{He}^{\textrm{Recfast}}) \epsilon^{\textrm{FOP}}. \label{eq:epsconv}
\ee

In our calculations, we adopt the energy injection rate convention of \cite{Furlanetto:2006wp} (equation [\ref{eq:epsFOP}]) and the helium fraction convention in RECFAST \cite{Seager:1999bc} (equation [\ref{eq:fHeRecfast}]), so for energy injection rates calculated outside of Recfast and fed into the Recfast evolution equations, we use the conversion $\epsilon \rightarrow (1+4 f_{He}^{\textrm{Recfast}})\epsilon$ to ensure consistency.


\end{document}